
%
%
%
\def\etal{\it et al. \rm}
\def\kms{km s$^{-1}$}

\def\atoms{atoms cm$^{-2}$}
\def\gsim{ \lower .75ex \hbox{$\sim$} \llap{\raise .27ex \hbox{$>$}} }
\def\lsim{ \lower .75ex \hbox{$\sim$} \llap{\raise .27ex \hbox{$<$}} }
\def\pp{\noindent\parshape 2 0truecm 16.0truecm 1truecm 15truecm}
\def\pp{\noindent\parshape 2 0truecm 15truecm 2truecm 13truecm}

\def\spose#1{\hbox to 0pt{#1\hss}}
\def\simlt{\mathrel{\spose{\lower 3pt\hbox{$\mathchar"218$}}
     \raise 2.0pt\hbox{$\mathchar"13C$}}}
\def\simgt{\mathrel{\spose{\lower 3pt\hbox{$\mathchar"218$}}
     \raise 2.0pt\hbox{$\mathchar"13E$}}}
\hrule height0pt
\magnification=\magstep1
\baselineskip 12pt
\parskip=6pt
\parindent=0pt

\hsize=6.5truein
\vsize=8.5truein


\font\titlefont=cmss17

\vskip 1.0truein

\centerline{\titlefont THE ORIGIN OF THE MAGELLANIC STREAM}

\vskip 1.5truein
\centerline{\titlefont Ben Moore and Marc Davis}
\vskip 0.4truein

\centerline{\it Department of Astronomy, University of California,
Berkeley, CA 94720, USA}

\hfil\eject

\vskip 0.5truein
\centerline {\bf ABSTRACT}
\vskip 8pt
\parindent=36pt
\vskip 0.4truecm

Recent measurements of the proper motion of the Large Magellanic Cloud (LMC)
show it to be moving in a nearly circular orbit around the Milky Way and to be
leading the Magellanic Stream which stretches in a well-confined arc
100$^\circ$
behind it. We present numerical investigations designed to  critically test
models of the origin of the Magellanic Stream. The most developed model is the
tidal model which fails to reproduce several of its characteristic properties:
(i) High resolution numerical simulations of tidal stripping show that stripped
material from the LMC would retain the internal velocities of the LMC and would
define a thick plane surrounding the Milky Way, not the tightly confined wake
actually observed.
(ii) There is no leading stream as would be expected if it were tidally
produced.
(iii) The observed radial velocity along the Stream is inconsistent with the
observed orbital parameters of the Magellanic Clouds.
(iv) The uniform variation in column density along the Stream cannot be
reproduced by tidal forces.
(v) No stars have been observed within the Stream, but stars
should be tidally stripped from the LMC as easily as gas.

We suggest an alternative model for the origin of the Magellanic Stream which
can explain all of its observed features and dynamics, as well as provide a
strong constraint on the distribution of gas within the halo of the Milky Way.
We propose that the Stream consists of material which was ram-pressure stripped
from the Magellanic System during its last passage through an extended ionized
disk of the Galaxy. This collision took place some 500 million years ago at a
galacto-centric distance of about 65 kpc, and swept $\sim 20$\% of the least
bound HI into the Stream. The gas with the lowest column density lost the most
orbital angular momentum. At the present time this material is at the tip of
the
Stream and has fallen to a distance of $\sim 20$ kpc from the Milky Way
attaining a negative velocity of 200 \kms. To prevent the stripped material
from
leading the Magellanic Clouds and attaining too large an infall velocity, we
postulate the existence of an extended dilute halo of diffuse ionized gas
surrounding the Milky Way. If the halo gas is at the virial temperature of the
potential well of the Milky Way ($2.2 \times 10^{6\ \circ}$K), its thermal
emission would contribute $\sim$ 40\% of the observed diffuse background
radiation in the 0.5-1.0 keV (M) band which is consistent with recent ROSAT
measurements as well as pulsar dispersion measures.  Ram pressure stripping
from
this extended disk and gaseous halo would explain the absence of gas in
globular
clusters and dwarf spheroidal companions to the Milky Way. Some fraction of the
observed high velocity clouds might be the infalling debris from previous
orbits
of the LMC through the extended disk.

\vfill\eject

\vskip 20pt
\centerline {\bf \S 1. INTRODUCTION}
\vskip 5pt
\parindent=36pt

The Magellanic Stream (MS) is a thin tail of neutral hydrogen (HI) which
stretches over $100^\circ$ across the sky and joins onto the Magellanic Clouds
(MCs). The MS is essentially continuous throughout its length with width $\sim
10^\circ$, although Mathewson \etal (1977,1984) identify six subclumps within
the Stream which they name MSI \--- MSVI. The MS has several distinguishing
features which allow a detailed comparison with models of its origin: The mean
column density of gas decreases the further away from the MCs, from $2\times
10^{20}$ \atoms\ close to the MCs, to $1\times 10^{19}$ \atoms\ at the tip of
the Stream. Furthermore, the radial velocity along the MS varies smoothly along
its length, from $-200$ \kms\ at its tip to $\sim 0$ \kms\ near the MCs. Table
1
summarises the main features of MSI \--- MSVI. Columns 2 to 4 list the mean
column density of neutral hydrogen, the angular distance of each cloud behind
the Small Magellanic Cloud (SMC), and the observed radial velocity calculated
assuming a circular velocity at the sun of 220 \kms. For continuity we have
also
shown the corresponding values for the SMC.

Several mechanisms for the origin of the MS have been proposed, although the
only model to be developed in detail is the tidal model proposed by Lin \&
Lynden-Bell (1979, 1982) and Murai \& Fujimoto (1980). These authors attempt to
explain the variation of radial velocity along the MS by assuming that it is
tidal debris torn from the MCs some 2 billion years ago when they were at
pericenter one full orbit ago. In order to reproduce the high negative velocity
at the tip of the MS, these models require a high transverse motion for the MCs
and an extended massive halo surrounding the Milky Way. Dynamical evidence of
large halos surrounding galaxies has been steadily accumulating. In particular,
orbital analyses of the satellite galaxies of the Milky Way and our motion
towards Andromeda suggest that our own halo extends to at least 200 kpc
(Zaritsky \etal 1989, Kahn \& Woltjer 1959).

The transverse motion of the Large Magellanic Cloud (LMC) has recently been
measured by Jones \etal (1991) and Tucholke \& Hiesgen (1991). These authors
have compared the present day positions of stars with old photographic plate
measurements to infer the proper motion of the LMC. Although the measurement
errors are fairly large, the results confirm the assumptions of previous models
that the MS is in fact trailing the MCs and lies close to the orbital plane.
Furthermore, the orbit of the LMC is observed to be approximately circular,
providing additional evidence for a massive halo extending beyond 50 kpc.
The original tidal models were based on the results from fairly crude numerical
simulations since the orbits of only a limited number of particles could be
followed. It is therefore of interest to examine these models in more detail,
taking advantage of the new orbital information and the ability to integrate
the
equations of motion of a large number of particles with high resolution.

In addition to the tidal scenario, several other models have been proposed for
explaining the origin of the MS (see Wayte 1991 for a review). These models are
not as well developed and all suffer from serious flaws. For example, Mathewson
(1987) proposed that the Stream consists of material swept from the intra-cloud
region of the MCs via ram pressure stripping by repeated collisions with high
velocity clouds in the halo. This idea is motivated by the proximity of several
such clouds to the MCs and MS, and the clumpy nature of the Stream. However,
several direct encounters are needed, all of which must occur in the same
direction and within the last half of a Gyr. This requires both a high level of
coincidence and a high space density of gas clouds in the halo. Furthermore,
this model would predict distinctly separated clumps of material within the
Stream and no correlation of column density with distance from the MCs.
Continuous ram-pressure stripping models, such as that of Liu (1992) combined
with a gravitational wake scenario, require a high density of gas at large
distances within the halo $\gsim 10^{-3}$ atoms cm$^{-3}$. These models can be
ruled out by the observed emission measure of low energy X-rays which is an
order of magnitude lower than this model would predict. Futhermore, these
models cannot explain the clumpy nature of the gas or the high
infall velocity at the tip of the Stream.

In Section 2 we perform high resolution numerical simulations of satellite
galaxies in order to investigate the dynamics of tidally produced features. We
use the results from these simulations to discuss several fundamental flaws in
the tidal model for the origin of the MS. In Section 3 we study possible orbits
of the Magellanic Clouds in the light of the uncertain transverse velocities of
the LMC and SMC. In Section 4 we propose and investigate a new model for the
origin of the MS and in Section 5 we discuss some interesting observational
consequences and predictions of this model. Our results
are concluded in Section 6.

\vskip 20pt
\centerline {\bf \S 2. TIDES AND TAILS}
\vskip 5pt
\parindent=36pt

In this section we examine the processes by which tidal forces remove material
from galaxies orbiting under the influence of an extended potential. In
particular, we want to study the effects of the Galaxy's gravitational field on
its satellite galaxies such as Fornax or the Magellanic Clouds. Our interest in
this study is primarily in the stripped material and the formation of tidal
features. (A complete analysis of the response of the internal structure of
satellites orbiting within an extended potential is in preparation.)

The tidal radius, $R_t$, specifies an approximate extent to a satellite galaxy
beyond which material will become bound to the
larger galaxy's deeper potential. For a bound
system of stars of total mass $M_s$ orbiting within an extended massive halo,
the tidal radius can be calculated approximately using
$$
\eqalignno{
R_t & \approx R_g (M_s/3M_g)^{1/3} \ ,
&(1)\cr}
$$
where $M_g$ is the mass within the pericentric distance $R_g$ and
the gradient $dM_g/dR_g$ across the satellite has been ignored.
Stars belonging to the satellite galaxy can gain energy via two body encounters
causing a slow but constant mass loss as stars escape through the tidal radius.
If the satellite is on an eccentric orbit then the rate of mass loss will
increase during its passage through pericenter where the tidal radius is
smallest.

Numerical simulations provide a convenient method of illustrating the forces at
work. First we construct a model satellite galaxy using a self consistent
density profile and stellar energy distribution. A King model provides a good
approximation to most of the satellites in the halo (King 1966).
Firstly, we shall model the Ursa Minor dwarf spheroidal
which has mass $\sim 10^7M_\odot$, core radius $\sim 150$ pc
and central velocity dispersion $\sim 8$ \kms. For this simulation
we will follow the evolution of 10,000 particles with the TREECODE described
by Barnes (1986). We use a timestep of half a million years, and with a force
resolution of 5 pc we conserve energy to better than 1\% over 5 Gyrs.

We model the Galaxy's halo by a fixed isothermal sphere, specified by a
constant
circular velocity $V_c=\sqrt{GM_g/R_g}=220$ \kms; hence the total mass
increases
linearly with distance to beyond the orbit of the satellite. For a system of
this mass the dynamical friction timescale is much longer than the orbital
period and can be ignored. We set the satellite in an
eccentric orbit with pericenter 20 kpc and apocenter 70 kpc, {\it i.e.} the
eccentricity, $e=0.6$. Figure 1a shows the system after 200,000 timesteps
$\equiv$ 10 Gyrs from a perspective looking down on the orbital plane. The
solid
line shows the orbital path taken by the satellite. Roughly 50\% of the mass
has
been lost over this timescale and lies within two long tails which both contain
similar amounts of material. The tail primarily below the
satellite is leading the satellite and the tail above is trailing.

The origin of the material in each tail is distinct but counter-intuitive. The
satellite becomes tidally extended along one axis which points towards the
center of the deeper potential, and is compressed along the perpendicular axes.
Material escapes primarily from both ends of the extended axis where the escape
velocity is lowest. The stripped material has a velocity equal to that of the
satellite's plus a residual velocity $\Delta v$. In this simulation, $\Delta v$
is small since it is some fraction of the internal velocity dispersion of the
satellite. Hence, material which was stripped $t$ Gyrs ago, presently lies at
the end of the tails, a distance $\sim t\Delta v$ kpc away from the satellite
where the units of $\Delta v$ are in \kms. Material which escapes from the
bottom of the satellite (the end of the extended axis pointing towards the
center of the potential) has less angular momentum and will orbit with a
slightly smaller major-axis. This implies that the orbital timescale will be
shorter than for the satellite and the stripped material will form a leading
tail. Similar reasoning applies to the formation of the trailing tail.
The width of the tails reflects both $\Delta v$ and the eccentricity of the
orbit. The velocity dispersion of stars within the tails is low, hence these
tidal features are very narrow. Figure 1b shows the simulation viewed in the
orbital plane. The stripped material lies in a very thin plane of comparable
width to the satellite.


This simulation unearths several serious problems when using the tidal model to
describe the origin of the Magellanic Stream. The 21cm
surveys have not detected a leading Stream, whereas we expect a similar amount
of material extending in front of the MCs. We also performed numerical
simulations of satellites on circular and highly eccentric orbits and in each
case both a leading and trailing tail formed. Furthermore, the stripped
material
lies within a highly uniform tail with very little variation in density along
its length, in conflict with the 21cm observations which show that the density
of material decreases by an order of magnitude along its length.

The above conclusions were drawn from a simulation of Ursa Minor. In order to
mimic tidal stripping from the MCs we now perform a simulation of a more
massive
system. Our model galaxy is again constructed with a King profile with unit
concentration, but with total mass $1.5\times 10^{10}M_\odot$. This model leads
to a central velocity dispersion of $\sim 50$ \kms, similar to the observed
dispersion within the LMC. The halo is modelled by a fixed potential, hence
the satellite in our model does not feel the effects of dynamical friction.
We treat this effect analytically by calculating the effective braking force
every 1000 timesteps. We take the mass of the satellite as the mass within the
observed tidal radius after fitting the profile to a King model. The
run is started with the satellite on an slightly eccentric orbit with
$e=0.3$ and from an initial galactocentric distance of 150 kpc.

Figure 2a shows the results of this calculation after 10 Gyrs when the
satellite
reaches a distance of about 50 kpc. The perspective in this diagram is viewed
from above the plane of the orbit, whereas Figure 2b is viewed from the orbital
plane. It is clear from these plots that the stars are escaping from the
model satellite with significant peculiar velocities which carry them well away
from the orbital plane within a Hubble time. The large $\Delta v$'s arise
because of the large internal velocity dispersion. These pictures clearly bear
little resemblance to the observed velocity and density structure of the
Stream.
It may be argued that these models are only simulating stars and not
a gaseous component. However, the gravitational tides affect all material
equally and we would expect to observe a significant number of stars within
the MS. Several groups have systematically searched for a stellar component
although no obvious enhancement of stars has been found in the vicinity of the
MS ({\it e.g.} Irwin 1991). As Lin \& Lynden-Bell originally
noted, these null detections present a further problem for the tidal model.

\vskip 20pt
\centerline {\bf \S 3. THE ORBITS OF THE MAGELLANIC CLOUDS}
\vskip 5pt
\parindent=36pt

Table 2 lists the observed orbital information of the MCs: $l$ and $b$ are the
galactic longitude and latitude; $R_g$ is the distance from the center of the
Galaxy; $v_r$ and $v_t$ are their radial and transverse velocities; the final
two columns list the mass of gas in neutral hydrogen and their total mass. Once
the transverse motions of the LMC and SMC are accurately measured, a model for
the mass distribution in the outer halo will completely specify the past and
future trajectories of these galaxies. The key unknown factor at present is the
transverse motion of the SMC; we can  however speculate on its value. The LMC
and SMC lie $\sim 20$ kpc apart in the sky, close enough for this to be an
unusual coincidence. They are within each other's tidal radii and  hence appear
marginally bound. Between the LMC and SMC lies $5\times 10^{8}M_\odot$ of HI
which has a smooth velocity gradient across its length. This bridge of gas also
contains many stars and young star clusters which suggests that it was formed
during a recent tidal encounter between the LMC and SMC. In order for material
to be stripped from the SMC, the two galaxies must have passed by each other
within a distance of about 5 kpc. Moreover, the encounter timescale must have
been relatively long, hence we infer that these galaxies are most probably
bound
together, although this is not prerequisite for our model discussed in Section
4.

If we impose the constraint that the SMC is gravitationally bound to the LMC,
then we find that this provides a tight constraint on its transverse velocity
$v_t$. From their present
positions, we can integrate the equations of motion backwards starting with
various values of $v_t$ to find values which keep the two galaxies
bound over a Hubble time. This calculation is straightforward once the mass
distribution in the halo is specified, and as in Section 2, we assume an
isothermal halo extending beyond the orbit of the MCs. Dynamical friction is
again included analytically, although we ignore mass loss through the tidal
radii of the satellites which would slowly decrease the rate of orbital decay.

Figure 3 shows two orbit calculations. In each plot the initial positions are
the present day locations of the MCs, the solid line shows the LMC's orbit and
the dashed line is that of the SMC. Figure 3a shows a bound orbit with
$v_t{_{(SMC)}}=210$ \kms\ and the observed value $v_t{_{(LMC)}}=235$ \kms.
Figure 3b also shows a bound orbit but with the transverse velocity of the LMC
equal to the $2\sigma$ upper limit from the observations $v_t{_{(LMC)}}=365$
\kms\ and we adopt $v_t{_{(SMC)}}=340$ \kms.
In both orbits the initial velocity vectors
are roughly parallel, and the MCs suffer two close collisions within the last
700 million years, at relative velocities $\sim 80$ \kms\ and separations
between 5 and 7 kpc. Extensive investigation of parameter space shows that in
order to maintain a stable bound orbit over at least half a Hubble time, the
transverse velocity vector of the SMC must align closely to that of the LMC and
have comparable magnitude. Differences in transverse velocity greater than 20\%
typically create unbound orbits within an orbital period. Orbits of higher
eccentricity have more freedom than circular orbits since the tidal field of
the
Galaxy does not play as important a role in unbinding the orbits.

{}From Section 2 we have seen that the velocity structure along a tidal tail
contains a great deal of information on the internal and orbital parameters of
the satellite galaxy. Observations of the radial velocity along such a feature
uniquely defines the orbit of the satellite. The tidal models for the MS as
proposed by Lin \& Lynden-Bell and Murai \& Fujimoto both require that the
transverse velocity of the LMC be significantly larger than the observed
circular velocity, in order to reproduce the high negative infall velocity at
the tip of the MS, a point several hundred million years and 100 kpc back along
the orbital path taken by the MCs.

Table 1 lists the galacto-centric radial velocity of the LMC
at a distance $\theta$ back along the orbital path at positions corresponding
to
the clouds MSI\---MSVI. Columns 5 and 7 give velocities corresponding to
the orbits in Figure 3a and Figure 3b respectively. These data should be
compared to the observations in column 4. At the tip of the MS, some
$110^\circ$ behind the LMC we expect an observed radial velocity of $-200$
\kms,
substantially larger than value predicted by integrating the observed orbit
back
in time. At the same point in the orbit of Figure 3a we would expect to observe
a radial velocity of only $-80$ \kms. Clearly, this cannot be the correct orbit
for the LMC within the context of the tidal model. Adopting a larger transverse
velocity for the LMC does not appear to alleviate this problem. In this case we
do observe large negative radial velocities along the orbital path:
at the tip of the Stream we expect a radial velocity of $-183$ \kms.
However, this orbit predicts that observed radial velocity should
{\it increase} to a maximum of $-210$ \kms\ between the location of MSIII and
MSIV, clearly in conflict with the observational data. For comparison with our
model proposed in the next section, in columns 6 and 8 we list the
galactocentric distances of the LMC at the same positions back along
the orbital path.

\vskip 20pt
\centerline {\bf \S 4. A NEW MODEL FOR THE ORIGIN OF THE MAGELLANIC STREAM}
\vskip 5pt
\parindent=36pt

\noindent{\bf 4.1 Ram pressure stripping by an extended disk}

Before proceeding with details, we shall outline the model we are proposing for
the origin of the MS. We assume that beyond the extent of the observed thin HI
disk of the Milky Way, hydrogen is still
present at a column density of $\lsim 10^{19}$ \atoms, but is mostly
ionized by the extragalactic background radiation. The MCs collided with this
extended disk of gas some 500 million years ago at a galactocentric distance
of about $65$ kpc. Material was ram-pressure stripped, mainly from the
intra-cloud region and the SMC, and began infalling towards the Galaxy.
The HI with lower column density lost more orbital angular momentum
than material of higher density, hence it has infallen the furthest into the
galactic potential and is moving with the largest negative velocity.
Material of column densities $\sim 10^{20}$ \atoms\
was barely perturbed and remains close to the MCs. A necessary component of the
model is that the Milky Way has a hot gaseous halo with density falling as
$R_g^{-2}$ beyond a core radius $\lsim 15$ kpc. This diffuse medium provides a
braking effect on the infalling material and prevents it from attaining
negative
velocities larger than those observed and allows the stripped material to fall
into the correct position in the sky.

In an attempt to determine how far the rotation curves of galaxies can be
resolved by observing the 21cm emission, van Gorkom \etal (1993) spent 100
hours
with the VLBA observing NGC 3198. This observation was capable of resolving
column densities of $5\times 10^{18}$ \atoms; however, at the observed column
density of $\sim 2\times 10^{19}$ \atoms, the flux dropped below
their detection limit in the interval
of one beamwidth (2.7 kpc). This sharp drop was also noted
by Corbelli \etal (1989) in their observation of M33. Motivated by these
observations and the pioneering work of Bochkarev and Sunyaev (1977), several
theoretical papers have subsequently argued that this sharp drop in column
density can be naturally explained by ionization of the HI by the cosmic
extragalactic UV radiation (Maloney 1992, 1993; Corbelli \etal 1993).
Recent Hubble Space Telescope (HST)
Faint Object Spectrograph observations toward
3C273 has shown that roughly half the observed Lyman alpha absorption clouds
have redshifts coincident with spiral galaxies near the line of sight but with
projected separations of up to 150 kpc, suggesting that disks extend in some
form, presumably mostly ionized, to this radius (Lanzetta \etal 1993).
In the context of standard models for
galaxy formation (White and Rees 1978), we would expect the formation of disks
to arise through the cooling of gas within an extended dark matter potential.
Applying standard cooling arguments to primordial gas within a massive dark
halo
we find that to distances of about $100$ kpc, the gas would have
had time to cool within a Hubble time.

At the solar distance, the column density of gas is $\sim 5\times 10^{20}$
\atoms. If the surface density falls as $R_g^{-1}$ beyond the solar radius then
we can expect a total column density, $N_{_H}^{tot}$, of order $10^{19}$
\atoms\
at a galacto-centric distance of 60 kpc. The self gravity of the gas would be
negligible and we can assume as did Maloney (1992) that the gas maintains a
Gaussian vertical distribution:
$$
\eqalignno{
n_{_H}(z) &= n_{_H}e^{{-z^2}/2\sigma_h^2} \ ,
&(2)\cr}
$$
with scale height $\sigma_h \approx R_g\sigma_{_{ZZ}}/V_c$ at large $R$.
At the solar radius the vertical velocity dispersion of the stars is
$\sigma_{_{ZZ}}\sim 15$ \kms. Observations of face-on galaxies show that
$\sigma_{_{ZZ}}$ is constant over a large range of radii, hence the midplane
density at large distances can be estimated from
$$
\eqalignno{
n_{_H}(0) &= {{{N_{_H}^{tot}}\over{\sigma_h\sqrt{(2\pi)}}} \ {\rm cm}^{-3}} \ .
&(3)\cr}
$$
Note that the resulting stripping is expected to be dependent on the total
column density, but not on $\sigma_{_{ZZ}}$ or the details of the vertical
density distribution in the disk. (We obtain indistinguishable results if we
approximate the disk as a very thin, uniform density slab.)

We can apply conservation of momentum to gain an insight into the
origin of the MS. As a first approximation consider the evolution of a gas
cloud
within the SMC of column density $N_{_H}$. If we approximate the disk as a very
thin slab, the velocity of the gas cloud after passing through the disk will be
simply $ v_i N_{_H}/(N_{_H}+N_{_H}^{tot})$, where $v_i$ is the
impact velocity.
{}From the orbit in Figure 3a, $v_i\sim 260$ \kms, hence the orbital velocity
of a
gas cloud of column density $10^{19}$ \atoms\ will suddenly be reduced to 130
\kms. This is significantly larger than the escape velocity of material within
the SMC which is $\approx 50$ \kms. A gas cloud of column density $10^{20}$
\atoms, will end up with a change in velocity of 24 \kms. Clouds
of this density in the MS must originate either from the outer edges of
the SMC and LMC or from within
the intra-cloud region where the escape velocity is very low.

There is an additional impulse which must be considered since the ionized disk
is rotating with circular velocity $V_c$. This impulse will be roughly
perpendicular to that previously described since the orbit of the MCs passes
close to the Galactic poles, and again we can apply
conservation of momentum to calculate the resulting velocity change. The net
effects are to increase the amount of stripped material and to force the gas to
infall along a spiral orbit which offsets the final distribution of MS
gas  from a great circle. This provides an additional constraint on $V_c
N_H^{tot}$ at the impact distance.

{}From this simple picture we can see how the main features of the MS arise.
Namely, the most dense clouds receive the smallest change in momentum and hence
will remain in orbit closest to the MCs. Likewise, the least dense clouds will
lose the most energy and fall furthest into the potential well of the Galaxy
and will achieve the highest radial velocity. In this model we would
expect a smooth transition of both density and velocity along the MS. The orbit
in Figure 3a results in an interaction between the MCs immediately before and
after they pass through the disk. Therefore gas of the same column density
might
be stripped from the LMC, SMC and the intra-cloud region, but with varying
amounts of momentum loss, hence the final distribution of material would be
both
continuous and clumpy.

To test this model in greater detail we follow numerically the orbits of gas
clouds of different mean column densities chosen to match clouds MSI\---MSVI.
As
well as the gravitational force we include an extra term to describe the
ram-pressure force, $\rho_m v^2$, where $\rho_m$ is the density of the medium
and $v$ is the velocity of the cloud. The key constraint on the model is the
position of the gas clouds in the sky and their radial velocity components. We
adopt the orbit shown in Figure 3a and place test particles of different column
densities at the position of the SMC 500 million years ago, the time when the
MCs were about to pass through the extended disk.
As expected, we find the resulting
momentum loss from the disk crossing to be sensitive only to the total column
density in the disk and not to the actual vertical density profile.

Figure 4a shows the subsequent evolution of these six clouds superimposed on
the
orbits of the MCs from Figure 3a. By the present time the clouds are actually
in
front of the LMC and SMC. This is obvious when one realises that because the
clouds lose energy, the apogalactic distance of their orbit decreases;
correspondingly, their orbital period decreases so they must overtake their
predecessor as projected on the sky. Clearly an additional force must play an
important role in the subsequent orbits of the stripped material.

\noindent{\bf 4.2 An extended gaseous halo for our Galaxy?}

We now consider the ram-pressure which could be provided by a large halo of hot
gas surrounding the Galaxy. Such a gaseous component is naturally expected as
material left over from the process of galaxy formation, as well as
intragalactic material which is continually accreting into the halo of the
Milky
Way. The gas is collisionally heated to the virial temperature of the dark
halo,
$2\times 10^6$K, thus allowing an extended spherical distribution about the
center of the halo. This halo of HII should not be confused with the hot
component of ionized gas which is localised within a few kpc of the plane of
the
Galaxy as in the standard 3 phase model of the ISM (McKee \& Ostriker 1977).

The orbit of the MS is sensitive only to the mass density of this halo, not
its temperature. Rather than a diffuse halo, the constraint below would equally
well be satisfied by a multiphase medium of lower pressure.
Direct evidence for the existence of such a diffuse halo
is the existence of a bow wave along the leading edge of the LMC where the
HI gradients are very steep (Mathewson \etal 1977).
In this case the ram pressure of the dilute gas is far too
weak to expell the gas from the LMC, but it does drive a shock
into the interstellar medium of the LMC.

We model the gaseous halo with a core radius $R_o$ and an $R_g^{-2}$ power law:
$$
\eqalignno{
n_{_{ISM}} &= n_o/[1 + (R_g/R_o)]^2 \ .
&(4)\cr}
$$
It is important to note that the orbits of the gas within the MS provides no
constraint on the shape of the gaseous halo within $R_g \sim 20$ kpc; we use
the parameterisation of equation (4) for convenience.

Figure 4b shows the orbits of the same six clouds after including the braking
effect of the above gaseous halo with $n_o=0.002$ atoms cm$^{-3}$ and
$R_o=12$ kpc. At a distance of 65 kpc this yields a gaseous halo of
$5\times 10^{-5}$ atoms cm$^{-3}$, and $\sim 7\times 10^{-4}$ atoms cm$^{-3}$
at the solar radius. The halo gas plays a crucial role in the
subsequent evolution of the stripped material and at the present time the six
test particles have fallen into roughly the same position as observed in
MSI\---MSVI. The final radial velocities and galactocentric distances of the
Clouds are tabulated in columns 9 and 10 respectively of Table 1. The agreement
with the observations is very good given the crude treatment of the dynamical
and hydro-dynamical processes involved. The only observation we have difficulty
in reproducing are the separations of the most dense clouds MSI\---MSIII, which
the model predicts to lie closer to the MCs. A more detailed simulation should
provide tighter constraints on the structure of the gaseous disk and halo.

We have a fair amount of freedom in our orbital parameters for the MCs, disk
column density and halo gas distribution. However, the above model provides
interesting constraints on all of these quantities. The final radial velocities
and galactic latitudes of the clouds are sensitive primarily to the halo gas
density and the mass of the dark halo within 65 kpc. If we increase $n_o$ by
more than 30\%, we find that the clouds can never attain the high infall
velocity of 200 \kms. If we decrease $n_o$ by more than 30\%, then both the
radial velocities and the positions in the sky become incompatible with the
data. Of course our models are not really sensitive to the density of
gas near the distance of the sun, since the MS lies outside this distance.
Thus $n_o$ and $R_0$ can be freely adjusted so long as the
density at $\sim 50$ kpc is held nearly constant.
The galactic longitude of the clouds are sensitive to both the total
column density and the rotational velocity of the disk. Figure 5 shows the
orbits of the test clouds from a perspective looking down upon the orbital
plane. Clearly the angular momentum provided by the rotation of the disk has
not
significantly perturbed the most dense clouds from their orbits on a great
circle. The least dense clouds follow a curved orbital path which brings them
back close to the great circle by the present time. In the next section we
discuss the constraints provided by the geometry of the MS.

\vskip 20pt
\centerline {\bf \S 5. FURTHER OBSERVATIONAL CONSEQUENCES}
\vskip 5pt
\parindent=36pt

\noindent{\bf 5.1 Magellanic Geometry}

The orbits of the satellite galaxies lie on great circles about the Galactic
center. In the case of the MS there are two effects which would spoil the
expected great circle: (i) The orbital plane is not being viewed from the
center
of the Galaxy, but from a position $\sim 8.5$ kpc away. (ii) The gas in the MS
receives a transverse velocity impulse due to the rotation of the disk. Figure
5
shows that this deviation is not large so long as the column density of disk
material is $\lsim 10^{19}$ \atoms. (This column density could be higher, but
the circular velocity at the impact distance would have to be reduced by the
same factor.)

If the tip of the MS had fallen close to the center of the galaxy then the MS
would appear curved in the sky due to our offset location. Figure 6 shows the
geometry: the large dashed circle is centered on the sun's position, S, and
extends 60 kpc to the position of the SMC at $l=300^\circ$ in the sun's frame.
The solid circle shows the perspective of an observer at the center of the
Galaxy, G, who would place the SMC at $l=288^\circ$. If material in the tip of
the Stream originated in the SMC, then it would form a great circle about G
represented by the dot-dash line, and would have an observed longitude in the
Galaxy's frame at point T where $l=108^\circ$. Converting this to the sun's
frame would place the tip of the Stream along the dotted line at
$l\sim 100^\circ$.

In our model, the tip of the Stream has actually fallen towards the Galaxy and
now lies at about $R_g\approx 25$ kpc. The tip therefore appears to lie along
the dashed line at T$^\prime$ which corresponds to a longitude of $87^\circ$,
agreeing well with the observed coordinates which are centered on $l=90^\circ$
$b=-40^\circ$. If the tip had fallen to within 20 kpc, then we would begin to
observe a strong curvature in the MS. The tidal model predicts that the tip of
the stream originated in the LMC and presently lies at a distance $\gsim 80$
kpc. Performing the same geometrical exercise, we find that in this model, the
tip of the Stream should appear to lie at $l\lsim 80^\circ$, 10 degrees away
from its observed position.

Radio maps of the neutral hydrogen distribution place $5\times 10^8M_\odot$ of
HI within the Stream where a mean distance of 50 kpc has typically been adopted
(Wakker \& Schwarz 1991). In our model, the bulk of the MS is closer to the
Galaxy, and the corresponding mass estimate is reduced, but only by about 25\%
since most of the high column density material is still at a distance of
about 50 kpc. The material in the MCs has mean metallicity somewhat below the
Galaxy's. During previous disk crossings, a similar mass of material
may have been swept from the MCs and has since accreted onto the Milky Way
enriching the disk and bulge material. It is possible that some of this
material
constitutes a large fraction of the high velocity clouds (HVCs) observed in
21cm
radio surveys. The highest density material stripped about a Gyr ago
would presently lie mainly
in the Southern hemisphere and have negative velocities of order
200\---300\kms. In addition, this picture would account for the below solar
abundances within the HVC's and their proximity to the sun. (Note that
dynamical friction decreases the distance to the LMC by about 10 kpc/Gyr,
therefore significant stripping can occur within the last couple of orbits.)
Wang (1993) discusses the possibility that tidally stripped material from
the MCs and dwarf galaxies could produce the bulk of the observed high
velocity clouds. A more detailed comparison between these models is in
preparation.

\noindent{\bf 5.2 The absence of gas in nearby satellites and globular
clusters}

Many globular clusters and several other less massive satellite galaxies orbit
within the halo of the Milky Way. It is something of a paradox that none of
these systems contain any neutral hydrogen. The absence of significant amounts
of gas in the globular clusters at galacto-centric distances less than 10 kpc
has been explained by ram pressure stripping as the clusters cross the observed
neutral hydrogen disk. However, continual stellar mass loss is expected to
deposit up to $150M_\odot$ of gas over a period of about 100 million years
(Tayler \& Wood 1975). Observational searches have not detected this material
and have placed upper limits of a few solar masses of HI
({\it i.e.} Smith \etal 1990).

The absence of gas within the clusters and satellites beyond the solar radius
provides an additional paradox since the disk is thought to extend to about 15
kpc. The most massive clusters {\it e.g.} $\omega$-Cen, have escape velocities
of the order of 40 \kms. The satellite galaxies in our halo are larger and more
massive, yet are less centrally concentrated and have escape velocities of
order
20 \kms. It is possible that an initial burst of supernovae could inject enough
energy into the ISM to clear the system of any gas left over from the formation
epoch. However, the satellites orbiting beyond the solar radius will have time
to form several thousand solar masses of hydrogen via stellar mass loss. A
natural consequence of our model is that the bulk of material left over from
star formation will be ram-pressure stripped after a single crossing of either
the neutral or ionized hydrogen disk. Any satellite crossing the disk at
distances less than 100 kpc will encounter a column density of hydrogen of at
least a few times $10^{18}$ \atoms, sufficient to strip all material of column
density $10^{19}$ \atoms. Furthermore, the diffuse halo provides a density of
$\sim 10^{-3}$ \atoms\ at the solar radius, sufficient pressure to continually
strip new material from the globular clusters as they orbit through the halo.
One prediction of this model is that we do not expect to observe low mass gas
rich satellite galaxies or globular clusters within a distance of about 100 kpc
from the center of spiral galaxies.

\noindent{\bf 5.3 The ionizing background}

The UV background responsible for ionizing the outer disk of the Milky Way will
also ionize the HI in the Magellanic Stream. For a uniform column density of
absorbing material, the timescale for ionization is $N_{_H}/\phi$, where $\phi$
is the incident ionizing photon flux at the Lyman limit. Detailed modelling of
the HI cutoff in the disk of NGC 3198 requires $5\times 10^3 \lsim \ \phi \
\lsim 5\times10^4$ photons cm$^{-2}$ s$^{-1}$. Observations of the proximity
effect at low redshift using the HST yield
background intensities consistent with this range (Kulkarni \& Fall 1993).

The column ionization timescale for gas at $N_{_H}=2\times 10^{19}$ \atoms\
would be at most 0.13 Gyrs, a timescale shorter than the half billion year
lifetime of the MS, therefore the ionization rate must be balanced by the
recombination rate which is $\propto n_e^2$. The width of the tip of the MS is
about 5 kpc assuming a distance of 25 kpc, giving a central density of atoms of
about $1.3\times 10^{-3}$ atoms cm$^{-3}$. The Hydrogen recombination
coefficient for gas at $10^4$K, $\alpha_{rec} \sim 2.6 \times 10^{-13}$
cm$^{-3}$ s$^{-1}$, therefore the column recombination timescale is $\approx 1/
(\alpha_{rec} n_e) \sim 0.1$ Gyrs. The similarity between these timescales
suggests that the recombination rate could balance the ionization rate. A more
detailed calculation would take into account the density structure of the
absorbing material, a diffuse component of radiation due to recombination
within
the gas etc. It is interesting to note that if the material in the tip of the
Stream was at a distance of $\sim 80$ kpc as the tidal model would predict,
then
the recombination timescale would be four times as long and the neutral
fraction
of the Stream would be small.

\noindent{\bf 5.4  Emission and dispersion measures}

Observations of soft X-rays by ROSAT provide a constraint on the amount of hot
gas in the halo. Burrows \& Mendenhall (1991) observed the shadowing effect of
the Draco cold interstellar gas cloud and conclude that a large fraction of the
low energy X-ray background arises from sources beyond 300 pc. Snowden \etal
(1993) used the shadowing effect to determine that essentially all of the M
(0.5-1.0 keV) band
radiation originates from distances beyond 60 pc. At least 40\% of the observed
background is due to discrete sources, such as distant QSO's, however, the
origin of a large fraction is still unaccounted for. Gas at the virial
temperature of the halo, $2\times 10^6$K, would primarily emit X-rays in the C
band $\approx 0.2$ keV, although a substantial component will also fall into
the M-band. Wang \& McCray (1993) analysed two deep ROSAT images
and concluded that up to 60\% of the M band radiation can be fit by discrete
sources, however, 40\% of the radiation in the M band arises in a diffuse
thermal component with a temperature of $\sim 2 \times 10^6$K.

In our model we require an extended halo of HII, rather than localised hot gas
close to the disk as in the standard three phase model of the ISM. The
integrated electron density along the line of sight quantifies the emission
measure, $E_m = \int n_e^2 dl$, of hot gas which could lie within the halo. We
can estimate $E_m$ for our adopted halo parameters by integrating the electron
density over the correct geometry. Wang \& McCray determine that the hard
diffuse component has an emission measure of $3.5\pm1.3\times 10^{-3}$
cm$^{-6}$
pc, in the direction $l=214^\circ$ and $b=-36^\circ$. Integrating over the
electron density in our model halo, in the same direction and using the
parameters adopted in equation (4), we find an identical emission measure as
observed by Wang \& McCray.

The emission measure toward any line of sight is obviously dominated by the
inner edge of our postulated halo, which is not constrained by the MS.
Furthermore, the cooling timescale for gas of temperature $2 \times 10^6$ K and
primordial abundance is $\approx 3 \times 10^9 n_{-3}^{-1}$ years, where
$n_{-3}$ is the density in units of $10^{-3}$ cm$^{-3}$. One expects the inner
edge of this postulated halo to have cooled off and to form a modest cooling
flow, unless reheating by cosmic rays or supernovae winds counteract the
cooling. Ignoring the contribution to $E_m$ from all the gas which could cool
within 10 Gyrs ($\sim 2R_\odot$) would reduce the emission measure by a factor
of two in the direction probed by Wang \& McCray. Note that the total mass of
ionized gas within the orbit of the LMC is about $3\times 10^9 M_\odot$,
roughly
0.5\% of the total mass within this radius. Of this amount, only
$10^8M_\odot$ lies within the core radius.

The dispersion measure of pulsars within the halo globular clusters provides
another measurement of the total column density of free electrons along the
line
of sight to the pulsar. Nordgren \etal (1992) show that the electron cloud of
the disk of the Milky Way extends to a scale height of between 500 kpc and 800
kpc, hence contributing a {\it minimum} dispersion measure,
$D_m = \int n_e dl\  \gsim $ 16 cm$^{-3}$ pc.
Within the Magellanic Clouds, the four observed pulsars yield
$D_m \sim$ 100 cm$^{-3}$ pc, with 40 cm$^{-3}$ pc arising from
our disk. Presumably, material in the MCs contributes a similar $D_m$, leaving
an excess of about 20 cm$^{-3}$ pc, which can be well explained by the
contribution from the diffuse halo in this direction which is 14 cm$^{-3}$ pc.
Another constraint arises from a pulsar within the globular cluster NGC 5024
which lies at high galactic latitude, 18.6 kpc above the disk and 18.9 kpc from
the sun, with $D_m = 25.0$ cm$^{-3}$ pc (Kulkarni \etal 1991).
The $D_m$ from the halo gas
towards this pulsar is $10\pm3$ cm$^{-3}$ pc, hence the total disk plus halo
$D_m$ agrees well with the observations.
The uniform slab model for the galactic disk
is a simple approximation to a complex medium which is filled with bubbles and
holes from supernovae winds. For example, the Galaxy is located within a local
void, relatively empty of ionized gas and with a scale of several hundred pc
(Phillips \& Clegg 1992). For this reason we consider the X-ray measurements to
provide the strongest constraint to our model at present.

\noindent{\bf 5.5 Confining the Magellanic Stream}

High resolution velocity maps of individual components of the stream show bulk
motions on small scales of the order 20 \--- 30 \kms. Velocities of this
magnitude would effectively double the size of the clouds in a few hundred
million years, hence, the initial column density of the MS would be higher by
at
least a factor of two which would be harder to strip from the MCs. Gas within
the MS of column density $\gsim 10^{20}$ \atoms\ is likely to be
gravitationally
bound. We must also consider the pressure, $P \propto n_e T$, provided by the
diffuse halo.  At the distance where the stripping occurred, the pressure
exerted by the halo is five times greater than the internal pressure of the MS,
where we adopted $10^4K$ as the temperature of the MS. This pressure imbalance
is at least an order of magnitude higher at the present distance of the tip of
the MS. However, it is likely that the components of the MS are in rough
pressure equilibrium with the diffuse halo gas; if they were not, the pressure
of a hot diffuse halo gas would crush them in a timescale of $10^7$ years. The
total pressures we are considering are very small and could easily be
supplemented by the pressure of a weak magnetic field of order $B_o \sim 1\mu
G$, which provides an internal pressure $B_o^2/(8\pi)$. Fields of this strength
are typical within the interstellar medium, and would naturally remain trapped
within the stripped material which forms the MS.

\vskip 20pt
\centerline {\bf \S 6. CONCLUSIONS}
\vskip 5pt
\parindent=36pt

The unusual velocity and density structure of the Magellanic Stream provide
strong constraints on models for its origin. We have investigated in some
detail
the tidal model and find several disparities between the predictions and the
observations. This leads us to suggest a new formation scenario for the MS in
which material is stripped from the Magellanic Clouds as it passes through an
extended ionized disk of the Milky Way. The subsequent orbits of this material
suggest that an additional ram pressure exerted by a diffuse halo gas, is
braking the infalling components of the MS. A simple investigation of this
model
suggests that it can account for all the observed features and dynamics of the
MS.

Observational and theoretical evidence is accumulating that the disks of spiral
galaxies do not end abruptly, as does the HI distribution, but rather continue
in a photoionized form beyond $\sim 15$ kpc. Hubble Space Telescope
observations
of the nearby quasar 3C273 show numerous Lyman alpha absorption lines at
coincident redshifts with spiral galaxies along the line of sight (Lanzetta
\etal 1993). This observation supports the notion that most spiral galaxies
have
extended disks of mostly ionized gas extending to scales of $\sim 100$ kpc. If
the Milky Way has such an extended disk, it partially strips gas out the
Magellanic clouds each time they pass through the galactic plane.

We find that reasonable parameters for the column density of the extended disk
and density structure of the diffuse halo results in a stream of stripped gas
which closely matches the observables of the Magellanic Stream, which stretches
in an arc for 100$^\circ$ behind the Magellanic clouds. We are able to match
the
velocity structure of the MS, the correlation of column density with angular
proximity to the MCs, the fact that the MS is neutral, and the apparent
trajectory of the MS on the sky. Most importantly, our model is fully
consistent with the observed transverse motion of the LMC.

A necessary component of our model is that a diffuse halo of gas surrounds the
Milky Way. This material provides an effective braking force on the stripped
material which plays a crucial role in determining the orbits of the gas. The
predicted X-ray emission from this ionized halo gas naturally accounts for
$\sim
40$\% of the background radiation in the 0.5\---1.0 keV band. The halo material
contributes to the dispersion measure of pulsars in globular clusters and the
MCs, however, the contribution from the disk is somewhat uncertain due to the
frothy nature of the local interstellar medium, hence does not provide as
useful
a constraint as the ROSAT measurements.

There are several observations which could directly measure the amount and
distribution of gas in the outer halos of spiral galaxies. HST observations of
nearby quasars, especially quasar-galaxy pairs, will provide the best
constraint
on the extent and distribution of hydrogen associated with spiral galaxies at
large galactocentric distances. X-ray observations of the halos of spiral
galaxies may be able to directly constrain hot ionized gas at the low expected
temperatures $\sim 0.2$ keV. High resolution neutral hydrogen observations of
gas rich satellite galaxies close to other massive spiral galaxies may be able
to detect stripped material. The velocity and density structure of these
features will enable tidal models to be distinguished from disk stripping
models. Finally high resolution ultraviolet spectroscopy toward extragalactic
sources using the Hubble Space Telescope will probe the physical conditions of
the diffuse halo gas of the Milky Way.

\vskip 0.5truecm

\vskip 20pt
{\noindent {\bf Acknowledgments}}
\vskip 5pt

We would like to thank Chris McKee for valuable comments and suggestions.
This research was supported by a NSF grant AST-9321540.

\vfil\eject

\vskip 20pt
{\noindent {\bf REFERENCES}}
\vskip 5pt

\pp Barnes J. 1986, {\it Nature}, {\bf 324}, 446.

\pp Bochkarev N.G. \& Sunyaev R.A. 1977,
{\it Astron. Zh.}, {\bf 54}, 957; trans. in Sov. Astro. {\bf 21}, 542.

\pp Burrows D.N. \& Mendenhall J.A. 1991, {\it Nature}, {\bf 351}, 629.

\pp Corbelli E., Schneider S.E. \& Salpeter E.E. 1989, {\it Astron. J.},
 {\bf 97}, 390.

\pp Corbelli E. \& Saltpeter E.E. 1993, {\it Ap.J.}, in press.

\pp Irwin M.J. 1991, {\it I.A.U. Symp. 148, The Magellanic Clouds},
ed. Haynes R. \& Milne D., 453.

\pp Jones B.F., Klemola A.R. \& Lin D.N.C. 1991, {\it Bull.Amer.Atron.Soc.}
{\bf 21}, 1107.

\pp Kahn F.D. \& Woltjer L. 1959, {\it Ap.J.}, {\bf 130}, 705.

\pp King I.R. 1966, {\it A.J.}, {\bf 71}, 64.

\pp Kulkarni S.R., Anderson S.B., Prince T.A. \& Wolszczan A. 1991,
{\it Nature}, {\bf 349}, 47.

\pp Kulkarni V.P. \& Fall S.M. 1993, {\it Ap.J.Lett.}, {\bf 413}, L63.

\pp Lanzetta K.M., Bowen D.V., Tytler D. \& Webb J.K. 1993, {\it Ap.J.}
{\it submitted}.

\pp Lin D.N.C. \& Lynden-Bell D., 1979, {\it M.N.R.A.S.}, {\bf 181}, 37.

\pp Lin D.N.C. \& Lynden-Bell D., 1982, {\it M.N.R.A.S.}, {\bf 198}, 707.

\pp Liu Y. 1991, {\it Astr.Astrophys.}, {\bf 257}, 505.

\pp Maloney P. 1992, {\it Ap.J.Lett.}, {\bf 398}, L89.

\pp Maloney P. 1993, {\it Ap.J.}, {\bf 414}, 41.

\pp Mathewson D.S., Schwarz M.P. \& Murray J.D. 1977, {\it Ap.J.Lett.},
{\bf 217}, L5.

\pp Mathewson D.S. \& Ford V.L. 1984, {\it I.A.U. Symp 108}, `Structure and
evolution of the Magellanic Clouds', ed. Van Den Bergh S. \& De Booer K.S.,
125.

\pp Mathewson D.S., Wayte S.R., Ford V.L. \& Ruan K. 1987, {\it Proc. Astr.
Soc. Australia.}, {\bf 7}, 19.


\pp McKee C.F. \& Ostriker J.P. 1977, {\it Ap.J.}, {\bf 218}, 148.

\pp Murai T. \& Fujimoto M., 1980, {\it P.A.S.J.}, {\bf 32}, 581.

\pp Nordgren T., Cordes J. \& Terzian Y. 1992, {\it A.J.}, {\bf 104}, 1465.

\pp Phillips J.A. \& Clegg A.W. 1992, {it Nature}, {\bf 360}, 137.

\pp Smith G.H., Wood P.R., Faulkner D.J. \& Wright D.E. 1990,
{\it Ap.J.}, {\bf 353}, 168.


\pp Snowden S.L., McCammon D. \& Sanders W.T. 1993, {\it Ap.J.}, in press.

\pp Tayler R.J. \& Wood P.R. 1975, {\it M.N.R.A.S.}, {\bf 171}, 467.

\pp Tucholke H.J. \& Hiesgen M. 1991, {\it I.A.U. 148}. `The Magellanic
Clouds',
ed. Haynes R. \& Milne D., 491.

\pp van Gorkom J.H., Cornwell T., van Albada T.S. \& Sancisi R. 1993,
in preparation.

\pp Wakker B.P. \& Schwarz U.J., 1991, {\it Astron.Astro.}, {\bf 250}, 484.

\pp Wang B. 1993, {\it Ap.J.}, {\bf 415}, 174.

\pp Wang Q.D. \& McCray R. 1993, {\it Ap.J.Lett.}, {\bf 409}, L37.

\pp Wayte S.R. 1991, {\it I.A.U. Symp. 148}. `The Magellanic Clouds',
ed. Haynes R. \& Milne D., 447.

\pp White S.D.M. \& Rees M.J. 1978, {\it M.N.R.A.S.}, {\bf 183}, 341.

\pp Zaritsky D., Olszewski E.W., Schommer R.A., Peterson R.C. \&
Aaronson M. 1989, {\it Ap.J.}, {\bf 345}, 759.

\vfil\eject

\vskip 20pt
{\noindent {\bf FIGURE CAPTIONS}
\vskip 5pt
\parindent=0pt

{\bf Figure 1.  } (a) A numerical simulation of the tidal effects of the Milky
Way on a low mass satellite such as Ursa Minor. The solid curve shows the
orbital path followed by the satellite which begins its 10 Gyr orbit at
coordinates (70,0). The viewpoint is looking down upon the orbital plane and
the
present position of the satellite is about (15,65) kpc. Roughly half of the
mass
has been stripped and has formed two long tidal tails. (b) The same simulation
viewed edge on shows that the stripped material occupies a very thin plane.

{\bf Figure 2.  } (a) A numerical simulation of the tidal effects of the Milky
Way on the LMC. After 10 Gyrs the tidally stripped material fills most of the
halo. The orbit of the satellite is discussed in the text. (b) The same
simulation viewed edge on shows that the stripped material occupies a thick
plane with most material concentrated inside the present orbit.

{\bf Figure 3.  } The past five billion years of the orbit of the LMC (solid
line) and SMC (dashed line). (a) We have used $v_t {_{(SMC)}}=210$ \kms\ and
$v_t {_{(LMC)}}=235$ \kms\ which allows a bound orbit over a Hubble time. (b)
As (a) except the predicted orbit of the MCs have been calculated adopting
transverse motions $v_t{_{(LMC)}}=365$\kms\ and $v_t{_{(SMC)}}=340$ \kms.

{\bf Figure 4.  } The simulated orbits of six gas clouds which represent MSI -
MSVI superimposed on top of our adopted orbits of the Magellanic Clouds (see
Figure 3a). (a) Gravity and ram pressure from the extended ionized disk are
considered. (b) as (a) and including the braking force from
an extended spherical halo of ionized gas.

{\bf Figure 5.  } The orbits of six gas clouds as described in Figure 4b, but
viewed from a perspective looking edge on to the orbital plane.

{\bf Figure 6.  } A sketch of the geometry of the Magellanic Stream. The
viewpoint is looking down from the North Galactic Pole. The large dashed circle
shows the sun's frame of reference for galactic coordinates centered on the
small open circle, whereas the solid circle shows the frame as viewed from the
center of the galaxy 8.5 kpc, away centered on the small filled circle. The
dash-dot line shows the great circle in which the SMC would orbit. Within the
plot we list the expected galactic longitude of the tip of the MS for different
values of its galacto-centric distance.

\bye